\input{aipcheck}

\documentclass[
    ,final            
  ]
  {aipproc}

\layoutstyle{8x11double}

\newcommand{\gsim}{\lower.7ex\hbox{$\;\stackrel{\textstyle>}{\sim}\;$}}
\newcommand{\lsim}{\lower.7ex\hbox{$\;\stackrel{\textstyle<}{\sim}\;$}}

\begin{document}

\title{Indirect Signatures of Gravitino Dark Matter}


\author{Alejandro Ibarra}{
  address={DESY,  Theory Group, Notkestrasse 85, D-22603 Hamburg, Germany},
  altaddress={Physik Department T30, Technische Universit\"at M\"unchen,\\
James-Franck-Strasse, 85748 Garching, Germany}
}

\begin{abstract}

Supersymmetric models provide very interesting scenarios
to account for the dark matter of the Universe.
In this talk we discuss scenarios with gravitino 
dark matter in $R$-parity breaking vacua, which
not only reproduce very naturally the observed dark
matter relic density, but also lead to a thermal history
of the Universe consistent with
the observed abundances of primordial elements and the observed 
matter-antimatter asymmetry. In this class of scenarios
the dark matter gravitinos are no longer stable, but
decay with very long lifetimes into Standard Model particles,
thus opening the possibility of their indirect detection.
We have computed the expected contribution from gravitino
decay to the primary cosmic rays and we have found that
a gravitino with a mass of $m_{3/2}\sim 150$ GeV
and a lifetime of $\tau_{3/2}\sim 10^{26}$ s could 
simultaneously explain the EGRET anomaly in the extragalactic
gamma-ray background and the HEAT excess in the positron fraction.

\end{abstract}

\maketitle


\section{Introduction}

A series of observations have provided in recent years compelling
evidences for the existence of dark matter in the 
Universe~\cite{Bertone:2004pz}.
These observations have revealed that the dark matter particle
has to be weakly interacting with the ordinary matter, 
long lived and slow moving (``cold'').
Among the most interesting candidates proposed 
stands the gravitino~\cite{Pagels:1981ke}
which is abundantly produced by thermal scatterings
in the very early Universe. If the gravitino is the lightest 
supersymmetric particle (LSP), a fraction of their initial population
will survive until today with a relic density which is calculable
in terms of very few parameters, the result being~\cite{Bolz:2000fu}
\begin{equation}
\Omega_{3/2} h^2\simeq 0.27
    \left(\frac{T_R}{10^{10}\,{\rm GeV}}\right)
    \left(\frac{100 \,{\rm GeV}}{m_{3/2}}\right)
    \left(\frac{m_{\widetilde g}}{1\,{\rm TeV}}\right)^2\;,
\label{relic-abundance}
\end{equation}
where $T_R$ is the reheating temperature of the Universe,
$m_{3/2}$ is the gravitino mass and $m_{\widetilde g}$ is the 
gluino mass.
In predicting the relic abundance of gravitinos,
the main uncertainty arises from our ignorance of the thermal 
history of the Universe before Big Bang nucleosynthesis (BBN) and
in particular of the reheating temperature after inflation.
However, we have strong indications that the Universe was very hot 
after inflation. Namely, the discovery of neutrino masses 
provided strong support to leptogenesis
as the explanation for the observed baryon asymmetry of
the Universe~\cite{Fukugita:1986hr}. This mechanism can reproduce the observed
baryon asymmetry very naturally if the reheating
temperature of the Universe was above $10^9$ GeV~\cite{bound}.
Therefore, the abundance of gravitinos can reproduce
the dark matter relic density
inferred by WMAP for the $\Lambda$CDM model,
$\Omega_{\rm CDM} h^2\simeq 0.1$~\cite{Spergel:2006hy},
for natural values of the input parameters in 
Eq.~(\ref{relic-abundance}).

Being capable of reproducing the correct relic density
is a necessary requirement for any dark matter candidate, 
but not the only one. Namely, the physical model accounting for
the dark matter should not spoil the successful predictions
of the standard cosmology. However, in scenarios with
gravitino dark matter, when $R$-parity is conserved, the  
next-to-LSP (NLSP) is typically present
during or after Big Bang nucleosynthesis,
jeopardizing the successful predictions
of the standard nucleosynthesis scenario. 
This is in fact
the case for the most likely candidates for the NLSP: the lightest
neutralino and the right-handed stau. More precisely, 
when the NLSP is the neutralino, the hadrons produced in the
neutralino decays typically dissociate the primordial 
elements~\cite{Kawasaki:2004qu},
yielding abundances in conflict with observations.
On the other hand, when the NLSP is a charged particle, $X^-$, the formation 
of the bound state $(^4{\rm He}\,X^-)$ catalyzes the production 
of $^6$Li~\cite{Pospelov:2006sc} leading to an overproduction 
of $^6$Li by a factor $300 - 600$~\cite{Hamaguchi:2007mp}.

Different solutions have been proposed to this problem.
For instance, in some specific supersymmetric models the NLSP can be a 
sneutrino~\cite{Kanzaki:2006hm} or a stop~\cite{DiazCruz:2007fc},
whose late decays do not substantially affect 
the predictions of Big Bang  nucleosynthesis.
Other solutions are to assume a large
left-right mixing of the stau mass 
eigenstates~\cite{left-right} or to assume some amount
of entropy production after NLSP decoupling, which dilutes the
NLSP abundance~\cite{Pradler:2006hh}. 
Our proposed solution consists in introducing a 
small amount of $R$-parity violation, so that the 
NLSP decays into two Standard Model particles 
before the onset of Big Bang nucleosynthesis, thus avoiding
the BBN constraints altogether~\cite{Buchmuller:2007ui}. 

\section{Gravitino Decay}

When $R$-parity is not exactly conserved, the gravitino LSP
is no longer stable. Nevertheless, the gravitino decay rate
is doubly suppressed by the Planck mass and by the
small $R$-parity violation~\cite{Takayama:2000uz}. 
Therefore, for the range of $R$-parity violating couplings
favored by cosmology, 
the gravitino lifetime ranges between $10^{23}$ and $10^{37}\;{\rm s}$
for $m_{3/2}=100\;{\rm GeV}$, which exceeds the age of the Universe 
by many orders of magnitude. Hence, even though the gravitino
is not absolutely stable, it is stable enough to constitute 
a viable candidate for the dark matter of the Universe, 
while preserving the attractive
features of  the standard Big Bang nucleosynthesis scenario 
and thermal leptogenesis. 

Interestingly, gravitinos could be decaying at a rate
sufficiently large to allow the detection of the decay products in
experiments, thus opening the possibility of the
indirect detection of the elusive gravitino dark matter.

In this talk we will discuss the possibility that gravitinos
are heavier than the gauge bosons. If this is the case,
gravitinos decay mainly into three different 
channels~\cite{Ibarra:2007wg,Ibarra:2008qg,Ishiwata:2008cu}:
$\psi_{3/2}\rightarrow \gamma \nu,\;  W^\pm \ell^\mp,\; Z^0 \nu$.
The fragmentation of the $W^\pm$ and the $Z^0$ eventually
produces a flux of stable particles, which we have simulated with the
event generator PYTHIA 6.4~\cite{Sjostrand:2006za}. 
On the other hand, the branching ratios of the different
decay channels are also calculable in the framework of supergravity
with broken $R$-parity, yielding a result that depends mainly 
on the gravitino mass (under the popular assumption of gaugino
mass unification at the Grand Unification scale). 

Dark matter gravitinos populate the halo with a distribution
that follows a density profile $\rho(\vec{r})$, where $\vec{r}$ denotes
the position with respect to the center of the Galaxy.
We will adopt for our numerical analysis the 
Navarro-Frenk-White density profile~\cite{Navarro:1995iw}:
\begin{equation}
\rho(r)=\frac{\rho_0}{(r/r_c)
[1+(r/r_c)]^2}\;,
\end{equation}
with $\rho_0\simeq 0.26\,{\rm GeV}/{\rm cm}^3$ and $r_c\simeq 20 ~\rm{kpc}$,
although our conclusions
are not very sensitive to the choice of the density profile.
Gravitinos at $\vec{r}$ eventually decay with lifetime $\tau_{3/2}$
producing photons, positrons, antiprotons and neutrinos at a rate per 
unit energy and unit volume given by
\begin{equation}
Q_x(E,\vec{r})=\frac{\rho(\vec{r})}{m_{3/2}\tau_{3/2}}\frac{dN_x}{dE}\;,
\label{source-term}
\end{equation}
where $x=\gamma, \; e^+, \; \bar p, \nu$ and $dN_x/dE$ is the energy 
spectrum of the particle $x$ produced in the decay. Remarkably,
the source function depends essentially on two unknown parameters,
namely the gravitino mass and the gravitino lifetime, yielding
a fairly predictive scenario. 

We will present in this talk the results for the gamma-ray 
flux~\cite{Ibarra:2007wg} and for the positron
and antiproton fluxes~\cite{Ibarra:2008qg}. The results
for the neutrino flux will be presented elsewhere~\cite{CGIT}.
The flux at Earth of the different particle species
are constrained by a series of experiments. EGRET measured
gamma-rays in the energy range between 30 MeV to 100 GeV. After 
subtracting the galactic foreground emission, the residual
flux was found to be roughly isotropic and thus attributed to 
extragalactic sources. An improved analysis
of the galactic foreground by Strong {\it et al.}~\cite{smr05},
optimized in order to reproduce the galactic emission,
shows a power law behavior between  50 MeV--2 GeV, but
a clear excess between 2--10 GeV, roughly the same
energy range where one would expect a signal from gravitino
decay. On the other hand, the flux of positrons 
has been measured by a series of experiments,
in particular by HEAT~\cite{Barwick:1997ig}. 
Although the measurements still suffer from large
uncertainties, it is intriguing that all the experiments seem to
point to an excess of positrons at energies larger than 
7 GeV, which is again the energy range where a contribution
to the flux from gravitino decay is expected. 
Lastly, the measurements of the antiproton flux by 
BESS~\cite{Orito:1999re} and other experiments do not show
any deviation from the predictions by conventional astrophysical 
models of spallation of cosmic rays on the Milky Way disk.
Future experiments such as the FGST (formerly GLAST)~\cite{FGST}, 
measuring the gamma-ray flux,
and PAMELA~\cite{Picozza:2006nm}, measuring antimatter fluxes,
will provide in the near future very precise measurements of the 
spectra of cosmic rays which will constitute decisive
tests of the decaying dark matter scenario.

\section{Gamma-ray Flux}

The gamma-ray flux from gravitino decay has two components.
The decay of gravitinos at cosmological distances will be detected
at Earth as a perfectly isotropic diffuse gamma-ray background
with a red-shifted energy spectrum. A second source of
gamma-rays is the decay of gravitinos in the Milky Way 
halo. For typical halo models, we find that the halo component 
dominates over the cosmological
one~\cite{Buchmuller:2007ui,Bertone:2007aw}, yielding
a slightly anisotropic gamma-ray flux.

\begin{figure}
  \includegraphics[width=6.8cm]{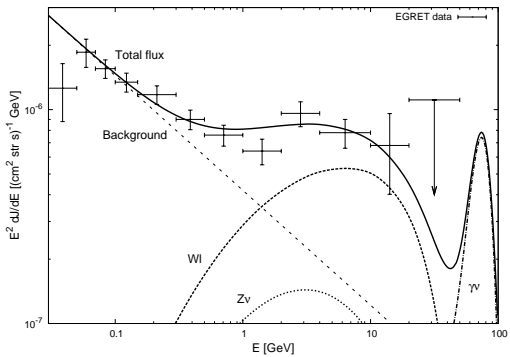}
  \caption{Contributions to the total gamma-ray flux
  for $m_{3/2}=150\,{\rm GeV}$ and $\tau_{3/2}\simeq 
  1.3\times 10^{26}{\rm s}$ compared to the EGRET data.
  In dotted lines we show
  the photon flux from the fragmentation of the $Z$ boson, in
  dashed lines from the fragmentation of the $W$ boson,
  and in dot-dashed lines from the two body decay 
  $\psi_{3/2}\rightarrow \gamma \nu$. The background is shown
  as a long dashed line, while the total flux received is shown as a
  thick solid line. 
\label{fig:flux}}
\end{figure}

The different contributions to the total gamma-ray
flux from gravitino decay are shown in Fig.~\ref{fig:flux} 
for a mass of $m_{3/2}\sim 150~{\rm GeV}$ and a lifetime 
of $\tau_{3/2}\sim  1.3\times 10^{26}~{\rm s}$~\cite{Ibarra:2007wg}.
To compare our results with 
the EGRET data~\cite{smr05}, also shown in the figure, we have averaged
the halo signal over the whole sky excluding a band of $\pm 10^\circ$
around the Galactic disk, and we have used an energy resolution 
of 15\%, as quoted by the EGRET collaboration in this energy range. 
The energy resolution of the detector is particularly
important to determine the width and the height of the monochromatic
line stemming from the two body decay $\psi_{3/2}\rightarrow \gamma \nu$.
The three contributions are dominated
by the halo component, the extragalactic component being
smaller by a factor of 2--3. Finally, we have adopted an 
energy spectrum for the extragalactic background 
described by the power law
$\left[\frac{d\Phi_\gamma}{dE}\right]_{bg}=4\times 10^{-7}
\left(\frac{E}{\rm GeV}\right)^{-2.5}
({\rm cm}^2 {\rm str}~{\rm s}~{\rm GeV})^{-1}$,
in order to provide a good agreement of the total
flux received with the data.

The predicted energy spectrum shows two qualitatively
different features. 
At energies between 1--10 GeV, we expect a continuous spectrum
of photons coming from the fragmentation of the gauge bosons.
As a result, the predicted spectrum shows a departure from the 
power law in this energy range that might be part of the 
apparent excess inferred from the EGRET data
by Strong {\it et al.}~\cite{smr05}.
In addition to the continuous component, the energy spectrum
shows a relatively intense monochromatic line at 
higher energies arising from the decay channel
$\psi_{3/2}\rightarrow \gamma \nu$. 

Even though the gamma-ray flux from gravitino decay
is expected to be anisotropic, we find that with these parameters the 
flux resembles an 
isotropic extragalactic flux in EGRET.
Namely, in the energy range 0.1-10 GeV, 
the anisotropy between the Inner Galaxy region
($|b|>10^\circ, 270^\circ\leq l \leq 90^\circ$)
and the Outer Galaxy region 
($|b|>10^\circ, 90^\circ\leq l \leq 270^\circ$)
is just a 6\%, well compatible with the EGRET data~\cite{smr05}.

\section{Antimatter Flux}

Antimatter propagation in the Milky Way is commonly described by
a stationary two-zone diffusion model with cylindrical boundary 
conditions~\cite{ACR}. 
Neglecting reacceleration effects
and non-annihilating interactions of antimatter in the Galactic disk,
the propagation can be described
by only a few parameters, which can be determined
from the flux measurements of other cosmic ray species, mainly from 
the Boron to Carbon (B/C) ratio~\cite{Maurin:2001sj}. 

The solution of the transport equation at the Solar System 
can be formally expressed by the convolution
\begin{equation}
f(T)=\frac{1}{m_{3/2} \tau_{3/2}}
\int_0^{T{\rm max}}dT^\prime G(T,T^\prime) 
 \frac{dN(T^\prime)}{dT^\prime}\;,
\label{solution}
\end{equation}
where $T$ is the kinetic energy and 
$T_{\rm max}=m_{3/2}$ for the case of the positrons 
while $T_{\rm max}=m_{3/2}-m_p$ for the antiprotons.
The solution is thus factorized into two parts.
The first part, given by the Green's function $G(T,T^\prime)$,
encodes all of the information about the astrophysics 
(such as the details of the halo profile and the 
complicated propagation of antiparticles in the Galaxy) 
and is universal for any decaying dark matter candidate. The
remaining part depends exclusively on the nature and properties
of the decaying dark matter candidate, namely the mass, the lifetime 
and the energy spectrum of antiparticles produced in the decay.
Finally, the flux of primary antiparticles at the Solar System
from dark matter decay is given by:
\begin{equation}
\Phi^{\rm{prim}}(T) = \frac{v}{4 \pi} f(T),
\label{flux}
\end{equation}
where $v$ is the velocity of the antimatter particle.

Clearly, if gravitino decay is the explanation for the extragalactic
EGRET anomaly, our predicted positron and antiproton fluxes
should not exceed the measured ones. 
In the scenario we are considering the gravitino mass and 
lifetime are constrained by requiring a qualitatively good
agreement of the predicted extragalactic gamma-ray flux with
the EGRET data: $m_{3/2}=150\,{\rm GeV}$ and 
$\tau_{3/2}=1.3\times 10^{26}\,{\rm s}$~\cite{Ibarra:2007wg}.
On the other hand,  the energy spectrum of antiparticles, $dN/dT$,
is determined by the physics of fragmentation.
Therefore, the main uncertainty in the computation of the
antimatter fluxes stems from the determination of the 
Green's function, {\it i.e.} from the uncertainties in the propagation
parameters and the halo profile. We have found in our
analysis that the uncertainties
in the precise shape of the halo profile are not crucial for
the determination of the primary antimatter fluxes.
On the other hand, the uncertainties in the propagation parameters 
can substantially change the predictions for the antimatter fluxes, 
even by two orders of magnitude for the antiproton flux.

Positrons and antiprotons have different properties and their
respective transport equations can be approximated by 
different limits of the transport equation.
Let us discuss each case separately.

\vspace{-0.4cm}
\subsection{Positron flux}

In the case of positron propagation, galactic convention
and annihilations in the disk can be neglected in the transport 
equation. We have solved the transport equation and computed the
Green's function for three propagation models, denoted
as  M2, MED and M1, which are consistent with the B/C ratio
and which yield, respectively, the  minimum, median and maximal
positron flux~\cite{Delahaye:2007fr}. 
The Green's function from dark matter decay 
can be well approximated by the following
interpolating function~\cite{Ibarra:2008qg}:
\begin{equation}
G_{e^+}(T,T^\prime)\simeq\frac{10^{16}}{T^2}
e^{a+b(T^{\delta-1}-T^{\prime \delta-1})}
\theta(T^\prime-T)\,{\rm cm}^{-3}\,{\rm s}\;,
\label{interp-pos}
\end{equation}
where $T$ and $T^\prime$ are expressed in units of GeV.
The coefficients $a$ and $b$ for each propagation model
can be found in Table~\ref{tab:fit-positron} for the NFW profile.
This approximation works better than a 15-20\% over the whole range
of energies. It should be stressed that this parametrization
of the Green's function is valid for {\it any} decaying dark matter particle,
not just for the gravitino.

\begin{table}[t]
\begin{tabular}{ccc}
 \hline
model & $a$ & $b$ \\ 
\hline
M2 & $-0.9716$ & $-10.012$  \\
MED & $-1.0203$ & $-1.4493$  \\
M1 & $-0.9809$ & $-1.1456$  \\
 \hline
\end{tabular}
\caption{\label{tab:fit-positron}
Coefficients of the interpolating function Eq.~(\ref{interp-pos}) 
for the positron Green's function, assuming a NFW halo profile
and for the M2, MED and M1 propagation models defined
in~\cite{Delahaye:2007fr}.}
\end{table}

With the previous parametrization of the Green's function,
it is straightforward to compute the interstellar positron
flux using Eqs.~(\ref{solution})
and Eq.~(\ref{flux}). The total positron flux received
at Earth receives, in addition to the primary flux 
from gravitino decay, a secondary component originating
in the collision of primary protons and other nuclei on the
interstellar medium, which constitutes the background to
our signal.

To compare our predicted flux with the observations
we choose to show the positron fraction,
defined as the total positron flux divided by the 
total electron plus positron fluxes:
\begin{equation}
{\rm PF}(T) = \frac{\Phi_{e^+}^{\rm{prim}}(T) + \Phi_{e^+}^{\rm{sec}}(T)}
{\Phi_{e^+}^{\rm{prim}}(T) + \Phi_{e^+}^{\rm{sec}}(T) 
+ k \;  \Phi_{e^-}^{\rm{prim}}(T) + \Phi_{e^-}^{\rm{sec}}(T)},
\end{equation}
where following \cite{Baltz:1998xv,Baltz:2001ir} we have left 
the normalization of the primary electron flux, $k$,
as a free parameter to be fitted in order to match the 
observations of the positron fraction. 
For the flux of secondary positrons, and the primary
and secondary electrons we used the parametrizations 
obtained in \cite{Baltz:1998xv} from detailed computer 
simulations of cosmic ray propagation~\cite{Moskalenko:1997gh}.

\begin{figure}
  \includegraphics[width=6.8cm]{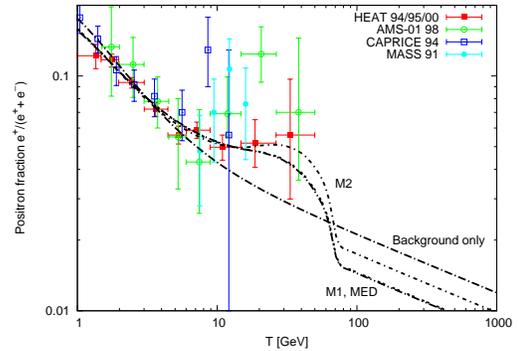}
  \caption{\label{fig:pos-frac}
  Positron fraction from the decay of gravitinos 
  for a NFW halo profile and the M2, MED and M1
  propagation models. Gravitino parameters are as
  in Fig.~\ref{fig:flux}.
  We also show
  for comparison the contribution to the positron
  fraction from spallation of cosmic rays on the
  Galactic disk, which constitutes the background to
  our signal.
  }
\end{figure}

We show in Fig.~\ref{fig:pos-frac} the positron fraction 
for different diffusion models in the case of the NFW profile
when $m_{3/2}\simeq 150\,{\rm GeV}$ and 
  $\tau_{3/2}\simeq 1.3\times 10^{26}\,{\rm s}$.
Interestingly, we find that gravitino parameters which predict a departure
from a simple power law in the extragalactic gamma-ray spectrum at energies
above 2 GeV (as observed by EGRET), {\it inevitably} predict a 
bump in the positron fraction at energies above 7 GeV (as observed
by HEAT)~\cite{Ibarra:2008qg}\footnote{The same conclusion 
has been independently reached by Ishiwata, Matsumoto and Moroi in
\cite{Ishiwata:2008cu}.}. Furthermore, the presence of 
this feature is not very sensitive to the many astrophysical uncertainties.
This remarkable result holds not only for 
the scenario of gravitino dark matter with broken $R$-parity,
but also for any other scenario of decaying dark matter with lifetime
$\sim 10^{26}\,{\rm s}$ which decays
predominantly into $Z^0$ and/or $W^\pm$ gauge bosons with
momentum $\sim 50\,{\rm GeV}$.

\vspace{-0.4cm}
\subsection{Antiproton flux}

The general antimatter transport equation can be simplified 
by taking into account that energy losses are negligible for 
antiprotons. Using this approximation, we have solved the
diffusion equation for three propagation models consistent
with the B/C ratio which yield the minimum (MIN),
median (MED) and maximal (MAX) antiproton flux~\cite{Delahaye:2007fr}.
We have found that 
the Green's function which describes antiproton propagation
from dark matter decay can be approximated by the following 
interpolating function~\cite{Ibarra:2008qg}:
\begin{equation}
G_{\bar p}(T,T^\prime)\simeq 10^{14}\,
e^{x +y \ln T +z \ln^2T}
\delta(T^\prime-T)\,{\rm cm}^{-3}\,{\rm s}\;,
\label{interp-antip}
\end{equation}
which, again, is valid for any decaying dark matter particle. 
The coefficients $x$, $y$ and $z$ for the NFW profile can be found in 
Table~\ref{tab:fit-antiproton}
for the different diffusion models.
In this case the approximation is accurate to a 5-10\%.
As in the case of the positrons, the dependence of the Green's
function on the halo model is fairly weak.

\begin{table}[t]
\begin{tabular}{cccc}
 \hline
model & $x$ & $y$ & $z$ \\ 
\hline
MIN & $-0.0537$& 0.7052 & $-0.1840$\\
MED & 1.8002 & 0.4099 & $-0.1343$\\ 
MAX & 3.3602 & $-0.1438$ & $-0.0403$ \\
 \hline
\end{tabular}
\caption{\label{tab:fit-antiproton}
Coefficients of the interpolating function Eq.~(\ref{interp-antip}) 
for the antiproton Green's function assuming a NFW halo profile
and for the MIN, MED and MAX propagation models 
defined in~\cite{Delahaye:2007fr}.}
\end{table}

The interstellar antiproton flux can be then straightforwardly
computed from Eqs.~(\ref{solution}) and Eq.~(\ref{flux})
using the previous parametrization of the Green's function
and the energy spectrum of antiprotons from gravitino decay,
$dN_{\bar p}/dT$.
However, this is not the antiproton flux measured by balloon
or satellite experiments, which is affected by solar modulation.
In the force field approximation~\cite{solar-modulation} 
the effect of solar modulation can be included
by applying the following simple formula that relates 
the antiproton flux at the top of the Earth's atmosphere and
the interstellar antiproton flux~\cite{perko}:
\begin{equation}
\Phi_{\bar p}^{\rm TOA}(T_{\rm TOA})=
\left(
\frac{2 m_p T_{\rm TOA}+T_{\rm TOA}^2}{2 m_p T_{\rm IS}+T_{\rm IS}^2}
\right)
\Phi_{\bar p}^{\rm IS}(T_{\rm IS}),
\end{equation}
where $T_{\rm IS}=T_{\rm TOA}+\phi_F$, with
$T_{\rm IS}$ and $T_{\rm TOA}$ being the antiproton kinetic energies 
at the heliospheric boundary and at the top of the Earth's atmosphere,
respectively, and $\phi_F$ the solar modulation parameter,
which we take $\phi_F=500$ MV.

\begin{figure}
  \includegraphics[width=6.8cm]{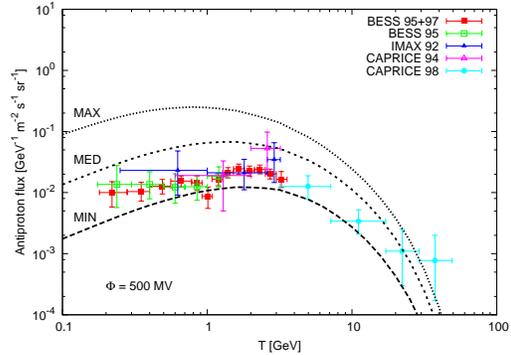}
  \caption{\label{fig:antiproton}
  Contribution to the antiproton flux at the top of the
  atmosphere from the decay of dark matter gravitinos
  for the NFW halo profile and the MIN, MED and MAX propagation
  models. Gravitino parameters are as in Fig.~\ref{fig:flux}.}
\end{figure}

We show in Fig.~\ref{fig:antiproton} the predicted antiproton
flux from gravitino decay for the MIN, MED and MAX diffusion models
when $m_{3/2}\simeq 150\,{\rm GeV}$ and 
  $\tau_{3/2}\simeq 1.3\times 10^{26}\,{\rm s}$.
From the plot, the extreme sensitivity of the 
primary antiproton flux to the choice of the diffusion model
is apparent: 
parameters that successfully reproduce the 
observed B/C ratio lead to antiproton fluxes that span over two
orders of magnitude. For a wide range of propagation parameters, the
total antiproton flux is well above the observations and thus
our scenario is most likely excluded. However, the MIN model yields
a primary flux that is below the measured flux and thus might be
compatible with observations.

\begin{figure}
  \includegraphics[width=5.5cm]{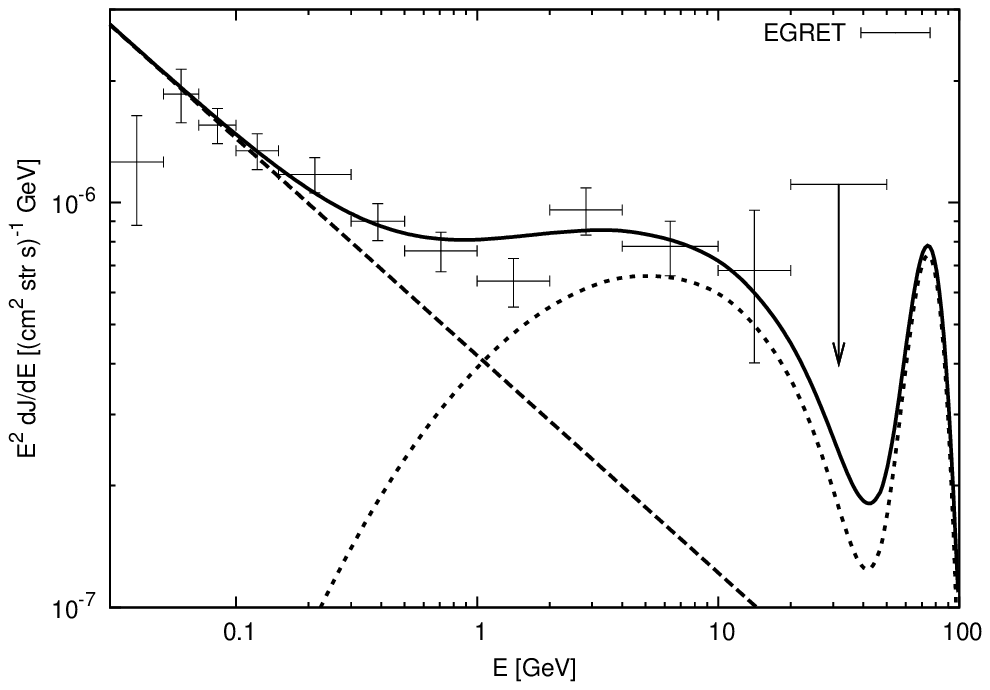}
  \includegraphics[width=5.5cm]{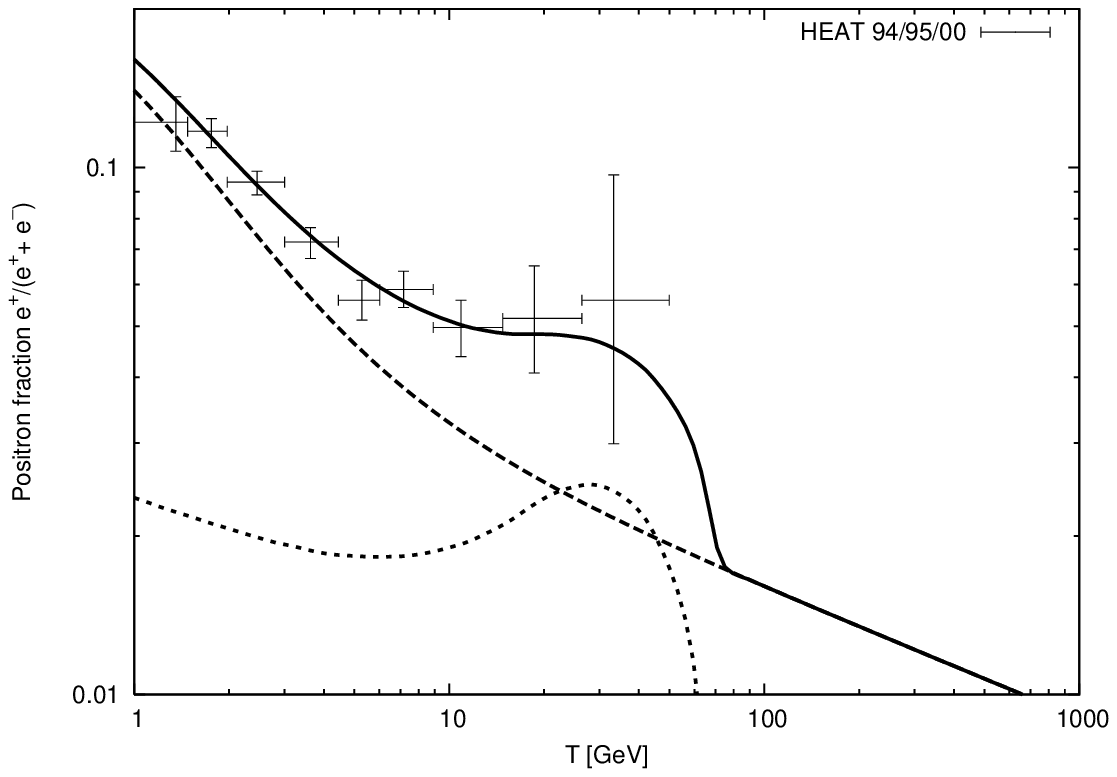} 
  \includegraphics[width=5.5cm]{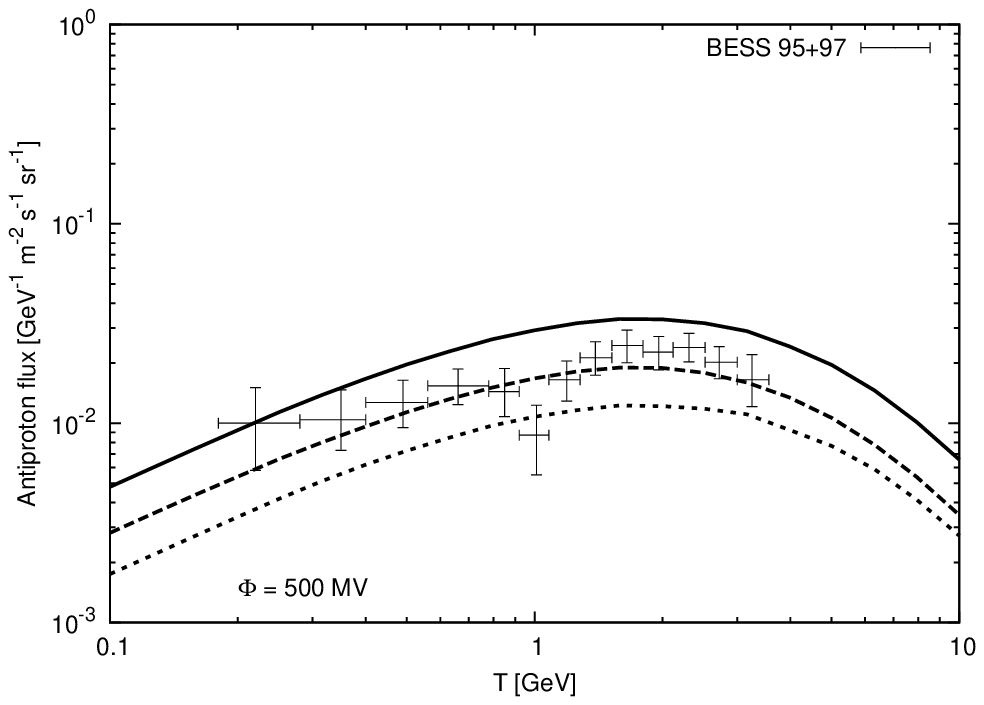}
  \caption{\label{fig:summary}\small
  Summary of the signatures of gravitino dark matter decay
  in the extragalactic gamma-ray flux (left), the positron fraction 
  (middle)
  and the antiproton flux (right), compared to the EGRET, HEAT and
  BESS data respectively.  The contribution from gravitino decay
  is shown with a dotted line, the brackground with a dashed line
  and the total flux with a solid line. 
  In these plots, we have adopted the MIN diffusion model, 
  $m_{3/2}\simeq 150\,{\rm GeV}$ and 
  $\tau_{3/2}\simeq 1.3\times 10^{26}\,{\rm s}$.
 }
\end{figure}

We have analyzed more carefully the predictions for the MIN model
computing the different contributions to the total antiproton flux. 
The result is shown in
the right panel of Fig~\ref{fig:summary}, where, for consistency,
we have adopted as background
the secondary antiproton flux calculated in \cite{Donato:2001sr} 
for the same MIN model. Although the primary antiproton flux is
smaller than the measured one, the total antiproton flux is a factor
of two above the observations. Nevertheless, in view of all the 
uncertainties that enter in the calculation of the antiproton flux,
it might be premature to conclusively rule out the scenario of
decaying gravitino dark matter. Namely, in addition to the uncertainties
stemming from degeneracies in the diffusion parameters, there
are also uncertainties from the nuclear cross sections and, to a lesser
extent, uncertainties from the description of the interstellar
medium and solar modulation (for a discussion of the various uncertainties
see \cite{Donato:2001sr}). Furthermore, 
we used a simplified diffusion model that neglects the effects of
reacceleration, energy losses and tertiary contributions. 
Therefore, there could be certain choices of parameters
or more refined diffusion models where the total antiproton flux is
consistent with experiments\footnote{Some works have reported
a deficit in the predicted secondary antiprotons compared to the observations 
and argued that this deficit could be connected with a contribution of 
primary antiprotons~\cite{Moskalenko:2001ya}.}.

\section{Conclusions}

The scenario of gravitino dark matter with broken
$R$-parity naturally reconciles three popular
paradigms that seem to be in mutual conflict:
supersymmetric dark matter, Big Bang Nucleosynthesis
and thermal leptogenesis. Moreover, the gravitino
decay products might become observable, thus
opening the possibility of the indirect detection
of the elusive gravitino dark matter.

We have shown that the EGRET anomaly in the extragalactic 
gamma-ray flux and the HEAT excess in the positron fraction 
can be simultaneously explained by the decay of a gravitino 
with a mass $m_{3/2}\sim 150$ GeV
and a lifetime of $\tau_{3/2}\sim 10^{26}$ s.
However,
the predicted antiproton flux tends to be too large, although 
the prediction suffers from large uncertainties and might be compatible 
with present observations for certain choices of propagation parameters.
Our results are summarized in Fig.~\ref{fig:summary}.

This remarkable result holds not only for 
the scenario of gravitino dark matter with broken $R$-parity,
but also for any other scenario of decaying dark matter with lifetime
$\sim 10^{26}\,{\rm s}$ which decays
predominantly into $Z^0$ and/or $W^\pm$ gauge bosons with
momentum $\sim 50\,{\rm GeV}$.

\vspace{0.1cm}
{\it Acknowledgements}
I would like to thank my collaborators G. Bertone, 
W. Buchm\"uller, L. Covi, M. Grefe, K. Hamaguchi,
T. T. Yanagida and especially D. Tran for a very pleasant and 
fruitful collaboration.

\end{document}